\documentclass[12pt]{article}
\usepackage{epsfig}

\parskip=\medskipamount 
plus.3em minus.5em \abovedisplayshortskip=.5em plus.2em minus.4em
\renewcommand{\thesection}{\arabic{section}}

\catcode`@=11

\@addtoreset{equation}{section} \@addtoreset{equation}{subsection}
\def\theequation{\ifnum\value{section}=0 \arabic{equation}\ignorespaces
\else \ifnum\value{section}=-1 A.\arabic{equation}\ignorespaces
\else \ifnum\value{subsection}=0
\thesection.\arabic{equation}\ignorespaces \else
\thesection.\arabic{subsection}.\arabic{equation}\ignorespaces
                             \fi
                        \fi
                   \fi}

{\catcode`\'=\active \def'{{}^\bgroup\prim@s}}

\catcode`@=12

\newcommand{\bq}{\begin{equation}}
\newcommand{\be}{\begin{equation}}
\newcommand{\fq}{\end{equation}}
\newcommand{\ee}{\end{equation}}
\newcommand{\bqr}{\begin{eqnarray}}
\newcommand{\beqs}{\begin{eqnarray}}
\newcommand{\fqr}{\end{eqnarray}}
\newcommand{\eeqs}{\end{eqnarray}}

\newcommand{\rf}[1]{(\ref{#1})}

\def\bop#1{\setbox0=\hbox{$#1M$}\mkern1.5mu
    \vbox{\hrule height0pt depth.04\ht0
    \hbox{\vrule width.04\ht0 height.9\ht0 \kern.9\ht0
    \vrule width.04\ht0}\hrule height.04\ht0}\mkern1.5mu}

\begin{document}
\thispagestyle{empty}

\begin{flushright}
\begin{tabular}{l}
hep-th/0503175 \\
\end{tabular}
\end{flushright}

\vskip .6in
\begin{center}

{\bf Geometric Solutions to Algebraic Systems}

\vskip .6in

{\bf Gordon Chalmers}
\\[5mm]

{e-mail: gordon@quartz.shango.com}

\vskip .5in minus .2in

{\bf Abstract}

\end{center}

A method to the explict solutions of general systems of algebraic
equations is presented via the metric form of affiliated K\"ahler 
manifolds.  The solutions to these systems arise
from sets of geodesic second order non-linear differential equations.
Algebraic equations in various fields such as integers and rational
numbers, as well as transcendental equations, are amenable.  The case 
of Fermat's set of equations is a subset.

\setcounter{page}{0}
\newpage
\setcounter{footnote}{0}

The solution to systems of algebraic equations is important to many
branches in mathematics.  Furthermore, their solutions over specified
fields such as integers and rationals are relevant to algebraic number
theory.  A primary point in this work is to bring together in a very
precise sense the solutions in areas of algebra and differential geometry.  
The role of modular invariance and automorphic forms is explicit in the
geometric solutions presented to the algebraic equations via the metric 
form of the Calabi-Yau manfolds.

A method is presented based on the geometry of Calabi-Yau manifolds,
their toric limits and non-compact relatives, to obtain the solutions
of polynomial and transcendental equations, in general.  The means is
obtained via the metric form on these manifolds and their geodesic flows,
or line integrals, in these spaces.  Given a set of equations, a complex 
manifold is constructed (e.g. toric variety, see for example 
\cite{TH}-\cite{EDM2}).  An initial point on the 
manifold is chosen which represents a solution to the algebraic system; 
subsequently, solving a set of second order non-linear partial differential 
equations generates the general solution set to the algebraic
system.  The primary complication, not explicitly given here,
is in solving these differential equations.  A uniform theory towards the
solution of algebraic systems is developed via the uniformization of
these toric (and potentially more complicated) examples.

The sets of algebraic equations of primary interest are those of the
form,

\bqr
 P_c(z_i) = \sum_{j=1}^m a_{\sigma(j,)}
 \prod_{l=1}^{n_{\sigma(j,l)}} z_{\sigma(j,l)}^{\tilde\sigma{(j,l)}}
  \ ,
\label{polynomial}
\fqr
with $c$ labeling the equation, and $\sigma_c(j,l)$ labeling the
$l$th term of the $c$th equation.  The numbers $a_{\sigma(j,)}$ are
coefficients multiplying the solved for variables $z_j$.  The
permutations of the terms are labeled by $\sigma$.  An example set is,

\bqr
  z_1^n + 2 z_2^m z_4^n + 3 z_3^n = 0
\fqr \bqr
  3 z_2^m + 2 z_3^n z_2^m  + z_4^m = 0  \ .
\fqr

The equations in \rf{polynomial} are general and contain, for
example, the well known case pertaining to Fermat's last theorem,

\bqr
P(z_i) = x^n + y^n - z^n
\label{fermatset}
\fqr
Equations with $m=\infty$, i.e. a power series, are also amenable by
the following approach.  These examples correspond to solutions to
transcendental equations.

The interest in this work is to find the solution set to the Diophantine
equations \rf{polynomial} over a general field.  The solutions
may be restricted to integers and rational numbers, such as the case
for the set in \rf{fermatset} in integers, and other examples.

Basically the methodology of the solution is given by the following
steps: (1) modeling a K\"ahler space via the polynomials $P_c(z_i)$
in terms of the holomorphic coordinates $z_i$ (and the anti-holomorphic
ones ${\bar z_i}$ in $P_c({\bar z}_i)$), (2) finding a simple sample
solution of (1) using a restricted set of coordinates, and then
(3) using a flow to other points in the manifold via a geodesic
equation.  Given the connectedness of the manifold all points are
connected by a line from the sample point to the solution of interest.

Pertaining to Fermat's set of equations, consider the point $x^n=1$
for $z=1$ in \rf{fermatset}, which has solutions $x=\exp(2\pi m/n$ for
$m=0$ to $m=n-1$.  The standard metric to the six dimensional manifold
modeled by the coordinates $(x,y,z)$ and $({\bar x},{\bar y},{\bar z})$
is constructed.  Then flows along geodesics are used to find the general
solution.  The obstacle is solving the non-linear equations (although
only second order) once the metric to the spaces are known.  However,
the solution is far more general than finding whether integer solutions
exist.

The spaces used in the work are primarily (toric) Calabi-Yau.
These manifolds have varieties defined by the equation set in
\rf{polynomial}, and may potentially be singular.  The latter
property is not irksome, as information is obtained from the
singularities describing the existence of solutions, away from
the locus of points in \rf{polynomial}.  Label the space pertinent
to \rf{polynomial} as $M_{P_c(z_i)}$ and its standard Riemannian
metric as $g_{\mu\nu}$.  Its K\"ahler so that both $g_{\mu\nu}=
\partial_{\mu}\partial_{\nu} \ln\phi(z_i,{\bar z}_i)$ (in terms
of $z$ and ${\bar z}$, $g=g_{i{\bar j}}$)  and its Christoffel
connection is derived as $\Gamma_{\rho,\mu\nu}= 1/2 \partial_\rho 
\partial_\rho \partial_\mu\ln \phi(z_i,{\bar z}_i)$ hold.

These polynomial sets of finite degree are modeled by a finite
dimensional Calabi-Yau space, the equation sets of infinite degree
(i.e. transcendental) are described by an infinite dimensional manifold.
The infinite dimensional manifold (or a manifold of large dimension)
actually comprises general solutions to the algebraic systems of lower
dimension by tuning their coefficients $a_i$ in \rf{polynomial}.
The K\"ahler geometrization of the algebra uniformizes to a large
extent the solution of the equations.

After finding the original point A to the solution of $P_c(z_i)=0$,
a geodesic flow equation is given from point A to point B.
The geodesic equation is, with the coordinates $x=(z_i,{\bar z}_i)$,

\bqr
 {d^2 x^\rho \over d\tau^2} + \Gamma^{\rho,\mu\nu}
    {d x_\mu\over d\tau} {d x_\nu\over d\tau} = 0
 \label{geodesic}
\fqr
with,

\bqr
\Gamma_{\rho,\mu\nu}= 1/2 \partial_\rho \partial_\mu\partial_\nu
\ln \phi(x_\mu,{\bar x}_\nu) \ .
\label{Chrisstofel}
\fqr
The coordinates $x$ contain both the holomorphic and anti-holomorphic
components describing the geometry.  Its complex form is

\bqr
{d\over d\tau} \left\{ {dz^i\over d\tau} +
  {dz_{\bar j} \over d\tau} \partial^{\bar j} \partial^i \ln(\phi)
    \right\}
 - {d^2 z_{\bar j}\over d\tau^2} \partial^{\bar j} \partial^i
        \ln(\phi)
  = 0 \ ,
\label{complexgeodesic}
\fqr
or

\bqr
{d\over d\tau} \left\{ {dz^i\over d\tau} +
  {dz_{\bar j} \over d\tau} g^{{\bar j},i} \right\}
 - {d^2 z_{\bar j}\over d\tau^2} g^{{\bar j},i}
  = 0 \ .
\fqr
These equations are second order in derivatives, but non-linear
because of the Kahler potential.

\begin{figure}
\begin{center}
\epsfxsize=12cm
\epsfysize=12cm
\epsfbox{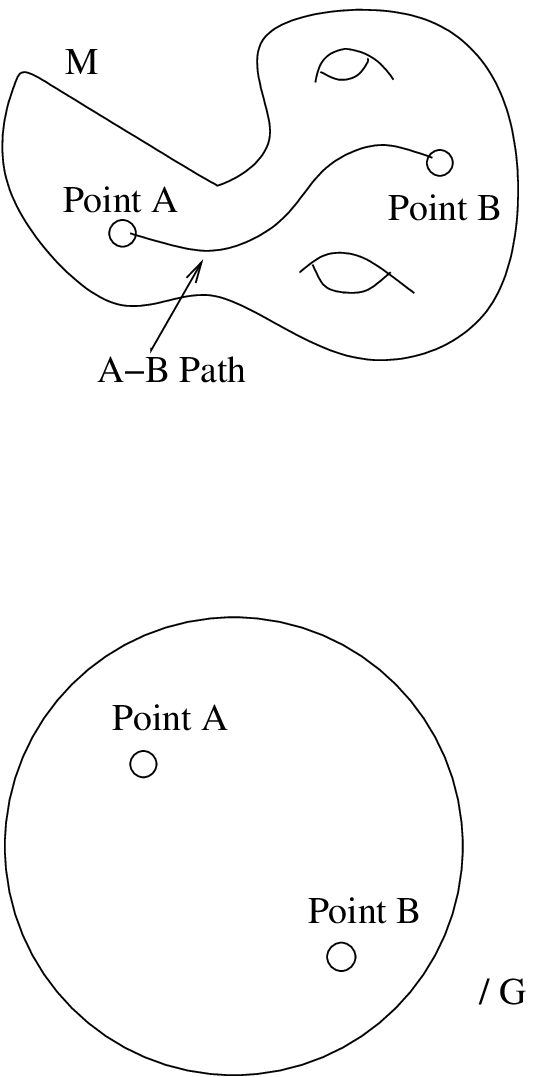}
\end{center}
\caption{An illustration of the path between two points A and B.}
\end{figure}

Given the initial coordinate A, the possible solution at point
B in \rf{polynomial} follows from the existence of the geodesic in
\rf{geodesic} or \rf{complexgeodesic}.  For example, the final
point may belong to the integers, rationals, primes, reals, or any
other set of interest.  If the solutions to the geodesic equations are
non-real, or singular, than the points are not allowed in the space
$M_{P_a(z_i)}$ spanning the original set of equations in
\rf{polynomial}.  In the well known Fermat's equations
for $n\geq 3$ the non-integral solutions could not be reached by a
geodesic and attempting to find one would result in a non-existent
solution to the non-linear pdes; for example an imaginary component
to the real coordinates $x_\mu$ might develop, or there might be a
singularity, signaling the failure of the integer solutions to exist
on the manifold.

A complication is in solving the coupled set of geodesic differential 
equations.  Solving these differential equations with the initial and 
final conditions, $x^i$ and $x^f$, is equivalent to solving for the 
algebraic solutions to $P_c(z_i)=0$.

In the toric Calabi-Yau context these differential equations are
uniformized in the sense that given the K\"ahler potential to the
infinite dimensional Calabi-Yau variety and a systematic solution
to the differential equations most algebraic systems could be solved
for after finding the geodesics.  This uniformization would correspond
to the set of polynomials of infinite degree with coefficients that
may be smoothly rotated to any value, hence spanning the basis of all
of the sets of polynomials.  This opens an avenue to finding relations
between solution spaces of seemingly disparate sets of algebraic systems.

The upshot of the analysis is that given,

\hskip .4 in 1) a metric on the variety parameterizing the set of
polynomial equations

\hskip .4 in 2) initial point A which is a simple solution

\noindent
then the existence of a point B, a number in a specified field, may
be found, or tested for, by solving a set of coupled second-order
non-linear partial differential equations (the geodesic equations) 
subject to the boundary conditions, i.e. points A and B.  In solving 
these differential equations, for example setting $x_f^\mu$ and 
$x_i^\mu$ to integers or rational numbers, one may systematically 
count and number the solutions.

We conclude with some discussion pertinent to the procedure of solving 
these algebraic systems.  

The set of algebraic equations is connected to orbifolded torii on the
geometric side via smooth interpolations of the potentially singular
(toric) Calabi-Yau manifolds.  These spaces have large volume limits
to torii quotients.  The sets of algebraic equations are thus connected
to the former spaces in various dimensions via these deformations
(including from one dimension to another by setting a coordinate to
zero).

The topology of the Calabi-Yau manifolds, and its cohomology and homology, 
and curve type and numbering,
should have an interesting interpretation in terms of the algebraic
equations and their solutions in relevant (possibly non-trivial)
fields.  Indeed, the existence and counting of specific solutions
to the algebraic equations, for example in the integer numbers, might be
related to these topological properties in a general sense.  There could 
be topological restrictions that specify classes of
algebraic systems that do not have integer solutions.

In the geometric solutions produced here, totally geodesic submanifolds 
or closed geodesics parameterize sets of 'special' sets of numbers 
(sub-fields) along these curves.  In a related point, the existence of 
these closed geodesics is often related to the countings in cohomological 
forms.

The algebraic systems are usually associated with Riemann surfaces of
varying genus, together with countings of their solutions 
in specified fields.  Modular invariance
appears both in these associated Riemann surfaces and affiliated sums,
for example in the generalized L-series.  A way of showing further
automorphic properties is by manifesting the metric properties of the
algebraic K\"ahler variety in the geodesic equation and its solution.

The metrics and the geodesic equation solutions are not evaluated in
this paper.  A differential form, via an operation ${\cal O}_{x_i}\cdot 
\ln \phi(z_i)=0$,
to find the K\"ahler potential, based on the polytopic definition
would be useful; in addition a classical solution to the D-terms of 
the ${\cal N}=2$ model with target the toric varieties would produce 
the utilized metrics.  The geodesic equations may be analyzed via 
a power series analysis in the proper time.  The number theoretic 
connection with the geometry is explicit and computable.  

\vfill\break

\end{document}